\documentclass[sigconf]{acmart}
\AtBeginDocument{%
  }

\setcopyright{acmlicensed}
\copyrightyear{2018}
\acmYear{2018}
\acmDOI{XXXXXXX.XXXXXXX}
\acmConference[Conference acronym 'XX]{Make sure to enter the correct
  conference title from your rights confirmation email}{June 03--05,
 2018}{Woodstock, NY}
\acmISBN{978-1-4503-XXXX-X/2018/06}

\usepackage[ruled,vlined]{algorithm2e}
\usepackage{listings}
\usepackage{xcolor}
\usepackage{booktabs}
\usepackage{tabularx}
\usepackage{graphicx}
\usepackage{array}
\usepackage{url}
\usepackage{xurl} 
\usepackage{pifont}
\newcommand{\OptBench}{\textsc{OptBench}\xspace}

\usepackage{tikz}
\newcommand*\circled[1]{\tikz[baseline=(char.base)]{
            \node[shape=circle,draw,inner sep=2pt] (char) {#1};}}

\lstdefinestyle{mystyle}{
  basicstyle=\ttfamily\small,
  numbers=left,
  numberstyle=\tiny,
  stepnumber=1,
  numbersep=8pt,
  keywordstyle=\color{blue},
  stringstyle=\color{teal},
  commentstyle=\color{gray},
  breaklines=true,
  frame=single,
  tabsize=2
}

\begin{document}

\title{\OptBench: An Interactive Workbench for  AI/ML-SQL Co-Optimization [Extended Demonstration Proposal]}

\author{Jaykumar Tandel}
\affiliation{%
  \institution{Arizona State University}
  \city{Tempe}
  \state{Arizona}
  \country{USA}}
  \email{jrtandel@asu.edu}

\author{Douglas Oscarson}
\affiliation{%
  \institution{Arizona State University}
  \city{Tempe}
  \state{Arizona}
  \country{USA}}
  \email{doscarso@asu.edu}

\author{Jia Zou}
\affiliation{%
  \institution{Arizona State University}
  \city{Tempe}
 \country{USA}}
\email{jia.zou@asu.edu}

%\affiliation{%
 % \institution{Inria Paris-Rocquencourt}
 % \city{Rocquencourt}
 % \country{France}
%}

\renewcommand{\shortauthors}{Trovato et al.}

\begin{abstract}
Database workloads are increasingly nesting artificial intelligence (AI) and machine learning (ML) pipelines and AI/ML model inferences with data processing, yielding hybrid ``SQL+AI/ML'' queries that mix relational operators with expensive, opaque AI/ML operators (often expressed as UDFs).
These workloads are challenging to optimize: (1) ML operators behave like black boxes, data-dependent effects (e.g., sparsity, selectivity, or cardinalities) can dominate runtime; (2) domain experts often rely on practical heuristics that are difficult to develop with monolithic optimizers; (3) introducing AI/ML operators will bring numerous co-optimization opportunities, such as factorization and pushdown, ML to SQL, linear algebra to relational algebra, etc., which will significantly enlarge the search space of equivalent execution plans and require new optimization strategies. At the same time, research prototypes for SQL+ML optimization are difficult to evaluate with fair, apples-to-apples comparisons, because each optimizer is typically developed on different platforms, and evaluated using different queries.

We present \OptBench, an interactive workbench for building and benchmarking query optimizers for hybrid ``SQL+AI/ML'' queries in a transparent, apples-to-apples manner. \OptBench runs all optimizers on a unified backend using DuckDB and exposes an interactive web interface that allows users to (i) construct query optimizers by leveraging and/or extending our abstracted logical plan rewrite actions; (ii) benchmark and compare different optimizer implementations over a suite of diverse queries while recording decision traces and latency; and (iii) visualize logical plans produced by different optimizers side-by-side. The demo illustrates how practitioners and researchers can prototype optimizer ideas, inspect plan transformations, and quantitatively compare optimizer designs on multi-modal inference queries within a single workbench.
Demonstration video can be found at : https://youtu.be/hHZAkUXhEWw
\end{abstract}

\maketitle

\begin{figure*} [h]
\centering
\includegraphics[width=1\textwidth]{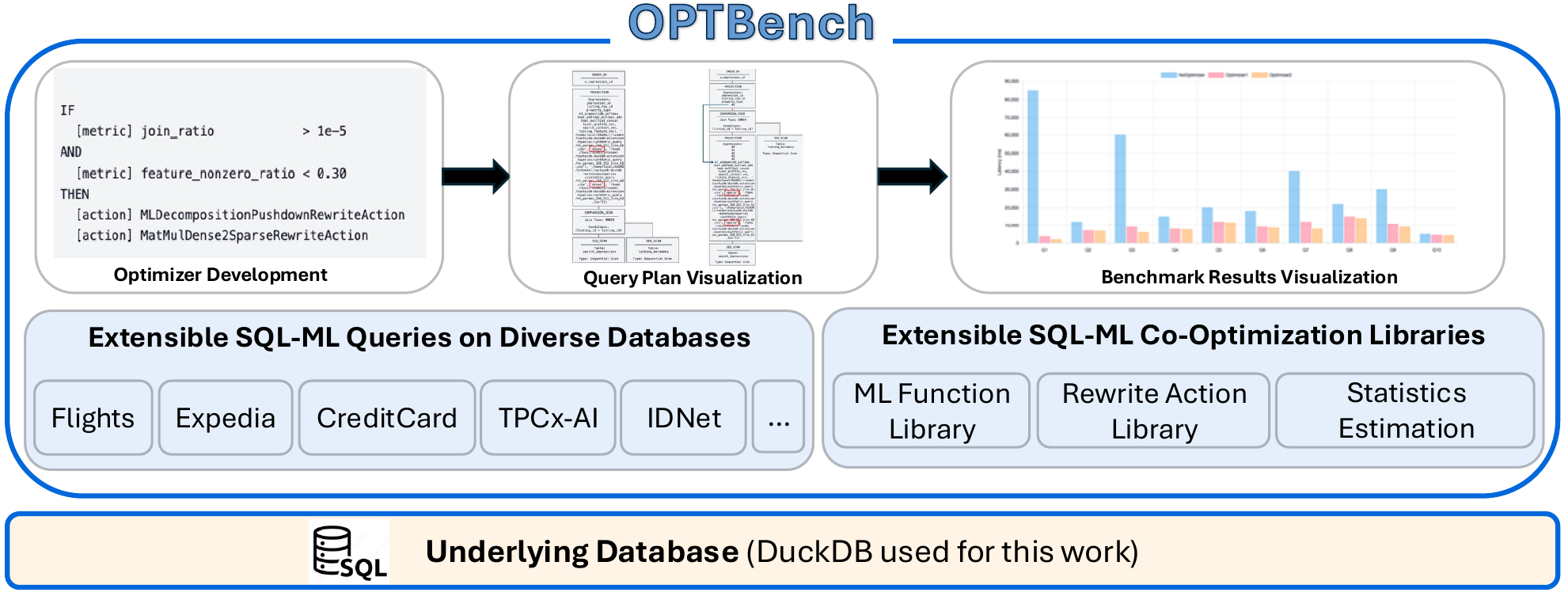}
\caption{\label{fig:overview} \small
OptiBench System Overview
}
%\vspace{-10pt}
\end{figure*}

\section{Introduction}
SQL engines are increasingly used as end-to-end execution substrates for data science and ML applications over relational data~\cite{park2022end, factorize-lmfao, chen2017towards, hellerstein2012madlib, zhang2025mitigating, gaussml, zhou2024serving, DBLP:journals/pvldb/ZhouCDMYZZ22, DBLP:journals/pvldb/YuanJZTBJ21, jankov2019declarative, zhou2026cactusdb, chowdhury2025inferf, guan2023comparison}.
In modern pipelines, feature extraction, data transformations, and even model inference are executed ``in-database'' alongside joins, filters, and
aggregations, producing SQL+ML queries that interleave relational operators with ML operators such as vector transformations, deep neural networks, convolutional neural networks, transformer models, and tree-based models.
While this integration can simplify deployment, reduce data movement~\cite{guan2023comparison}, and alleviate privacy issues~\cite{guan2025privacy}, it also introduces new performance pathologies. For example, a missed opportunity to exploit data sparsity, factorize an ML function, or convert a linear algebra operator to SQL, could dominate end-to-end latency.

Unfortunately, optimizing SQL+ML workloads remains challenging for several reasons.

First, it is hard to extend traditional query optimizers to support SQL-ML co-optimizations. On one hand, ML operators are often expressed as UDFs or compound expressions that appear opaque to traditional optimizers, which are primarily designed around
relational algebra and well-understood physical operators. In addition, effective heuristic rules, such as ``apply strong filters before invoking an expensive model'' or ``switch between dense and sparse
kernels depending on sparsity'' are hard to encode and validate inside a monolithic optimizer due to the lack of first-class extensibility and cost introspection for ML operators in traditional relational database systems. 

Second, research and prototype optimizers for SQL+ML are difficult to compare fairly: each is typically implemented inside a different system or codebase,
exposing different plan representations, statistics interfaces, and evaluation pipelines.  For example, MADLib~\cite{hellerstein2012madlib} was an extension of PostgreSQL, Raven~\cite{park2022end} was developed on SQLServer, LMFAO~\cite{factorize-lmfao} was implemented from scratch, Morpheus~\cite{chen2017towards} was built on R and Oracle Enterprise. IMBridge~\cite{zhang2025mitigating} was developed on OceanBase and DuckDB, and InferF~\cite{chowdhury2025inferf} was implemented on top of Velox. As a result, there is no easy way to compare the SQL-ML co-optimization techniques implemented in these systems in an apples-to-apples manner.

Third, performance is often governed by data-dependent and workload-dependent effects, and it requires a set of diverse workloads and datasets to evaluate different optimization techniques. However, different SQL-ML co-optimization systems adopt different workloads and datasets in their evaluation plans, without a standard. Moreover, the optimization process is often treated as a black box, where users see only final runtimes or costs, with little visibility into which decisions
were made, which rules fired, or how those decisions depended on data and workload statistics.

This demo paper addresses these gaps with \OptBench, an interactive workbench for transparent optimizer design and benchmarking on SQL and SQL+ML workloads.
\OptBench is motivated by three common user stories, where the users can be system builders, researchers, and data scientists:
\textbf{(1) SQL-ML optimizer development and performance evaluation.} A user designed a new SQL-ML co-optimization strategy (e.g., a new heuristic rule or a new optimal-plan search algorithm) for SQL+ML workloads and needs to fairly compare its efficiency and effectiveness against other optimization strategies, under controlled access to plan structure and statistics.
\textbf{(2) SQL-ML optimizer benchmark and comparison.} A user may want to compare different optimizer design decisions and algorithms on a set of common workloads with the same hardware resources and software configuration.
\textbf{(3) SQL-ML optimizer debugging and enhancement.} A user found that running some data science pipelines inside a relational database is unexpectedly slow, but lacks insight
into whether the issue stems from join ordering, premature ML evaluation, or incorrect cardinality assumptions. They need to modify the optimizer (e.g., to enable or disable some rules) and run it for debugging purposes.
\OptBench supports all of these scenarios by providing a unified and extensible environment, where optimizer can be constructed, inspected, and evaluated.

\OptBench is built on DuckDB, consisting of several key components: \textbf{(1) An extensible library of ML functions} encapsulated as DuckDB C++ UDFs, which can be used to compose complex ML inference workflows; \textbf{(2) An extensible library of query plan rewrite actions}, with each applying a certain SQL-ML co-optimization technique, converting the input query plan to an output query plan that will achieve equivalent query results but with different performance implications; \textbf{(3) An extensible library of statistics methods} for computing and caching data statistics, predicate selectivities, and intermediate result cardinalities; \textbf{(4) An extensible library of diverse SQL-ML queries} for evaluating different optimizers fairly; \textbf{(5) A web-based user interface} for developing, uploading, inspecting, and evaluating optimizers. The reason we select DuckDB as the foundation system of \OptBench is that it provides a compact, modern, and highly hackable optimizer and execution engine with strong support for UDFs and plan rewrites, enabling rapid prototyping and controlled benchmarking of SQL-ML optimizers. In addition, DuckDB’s in-process execution model, Python/R integration, and low deployment overhead make \OptBench directly accessible to data scientists, allowing them to experiment with in-database ML pipelines, inspect execution plans, and diagnose performance issues without setting up complex distributed systems. In the future, we will also extend the framework to other systems.

\paragraph{Contributions.}
This demo makes the following contributions:
(1) We propose and have prototyped \OptBench, an interactive optimizer workbench for easily developing new SQL-ML query optimizers,  executing SQL+ML queries with different optimizers for transparent inspection and apples-to-apples comparison;
(2) We provide a web-based user-interface for developing or uploading new optimizers, executing selected queries using two or more query optimizers with side-by-side query plan and performance results visualization. We also provide suite-level benchmarking that enables controlled, apples-to-apples comparison of heterogeneous optimizers; and
(3) We designed two different types of interactive demonstration scenarios: (i) Using \OptBench library or directly using the web interface to develop and upload new query optimizers; (ii) Using \OptBench user interface to select different queries and query optimizers for detailed performance comparison and analysis.

\section{System Overview}

As illustrated in Fig.~\ref{fig:overview}, \OptBench is built on DuckDB, consisting of the following components. 
%’s workflow. Given an input SQL or SQL+ML query, OptiBench captures the query’s logical plan and computes a set of optimizer-relevant metrics using a shared metric library and estimation pipeline. Users compose optimizers by writing metric-driven rules that trigger reusable rewrite actions (from an action library). These user-defined optimizers (e.g., Opt1/Opt2) coexist with baseline and advanced backends, such as DuckDB’s optimizer and a search-based MCTS optimizer. Finally, OptiBench benchmarks all selected optimizers over a query suite and provides visual tools to compare resulting plans (QP1 vs QP2) and performance on the query bench.

\subsection{Extensible ML Function Library}
\label{sec:ml-function-library}
\OptBench targets \emph{multimodal SQL+ML inference queries} that embed model inference execution inside SQL via user-defined functions (UDFs). To support this
setting in a self-contained way, our DuckDB-based prototype provides an extensible library of ML operators as scalar/table functions that can be used to compose 
SQL expressions. These functions cover common building blocks for inference pipelines, including linear algebra operators,
data preprocessing functions, neural network operators, and tree-based models. 

Table~\ref{tab:ml-functions} shows an example collection of ML functions supported in \OptBench and shipped with the demo. In our query suite, these functions are composed to form higher-level
inference operators (e.g., multi-layer neural networks expressed as nested linear algebra operators, or decision-forest inference expressed as aggregation of a collection of decision tree inferences). %which OptiBench then rewritesvia actions such as decomposition/pushdown, kernel selection, or relationalization.

\begin{table}[t]
\centering
\small
\caption{Example of ML functions used to express SQL+ML inference queries in the \OptBench demo.}
\setlength{\tabcolsep}{4pt}
\renewcommand{\arraystretch}{1.12}
\begin{tabularx}{0.5\textwidth}{@{}>{\raggedright\arraybackslash}p{0.44\columnwidth} X@{}}
\toprule
\textbf{Function} & \textbf{Role in SQL+ML queries} \\
\midrule
\texttt{matrix\_multiply} (\texttt{mat\_mul}) & Core linear operator used to express NN layers. \\
\texttt{matrix\_addition} (\texttt{mat\_add}) & Linear layer bias/addition and intermediate composition. \\
\texttt{conv2d} & 2-dimensional convolutional operator. \\
\texttt{softmax} & Normalization for classification/ranking logits. \\
\texttt{sigmoid} & Activation for binary/logistic inference. \\
\texttt{distance} & L1 or L2 distances between two vectors. \\
\texttt{cosine\_sim} & Cosine similarity between two vectors. \\
\texttt{argmax} & Prediction operator over class scores. \\
\texttt{torch\_dnn} & Fused neural-network inference operator (backend-supported). \\
\texttt{min\_max\_scaler} & Feature preprocessing / normalization. \\
\texttt{one\_hot\_encoder} & Categorical feature expansion. \\
\texttt{kmeans} & Clustering operator for customer segmentation / grouping. \\
\texttt{naive\_bayes} & Text/feature classification using Naive Bayes scoring. \\
\texttt{llm} & LLM-based inference call (black-box model call). \\
\texttt{decision\_tree} & Tree-model inference primitive. \\
\texttt{decision\_forest} & Ensemble inference (multiple trees) using XGBoost or Random Forest. \\
%\texttt{gpt} & Remote invocation of various OpenAI GPT APIs.\\
%\texttt{ml\_decision\_forest\_table} & Relational/table form of forest inference for query integration. \\
\bottomrule
\end{tabularx}
\label{tab:ml-functions}
\end{table}

\subsection{Extensible Rewrite Actions}
\label{sec:rewrite-actions}
\OptBench defines a \emph{Rewrite Action} as a reusable, composable transformation that rewrite a logical query plan by applying a SQL-ML co-optimization technique while ensuring the equivalence of query output given the same input.
These rewrite actions serve as the ``building blocks'' of SQL-ML query optimizers. A SQL-ML query optimizer attempts to select an optimal sequence of applying these actions to generate an improved query plan. Rule-based optimizers trigger actions when metric predicates hold,
while cost-based optimizers enumerate and score plans obtained by applying sequences of rewrite actions.
\OptBench ships with a small built-in library of reusable rewrite actions, with examples shown in Table~\ref{tab:action-library}. They can be
used to develop rule-based and cost-based optimizers.

%\jia{Jay, Please add explanations to those rewrite actions here. - Done}

\paragraph{Dense/Sparse kernel selection.}
This action targets matrix multiplication expressions such as \texttt{matrix\_multiply(X, W)} and selects between dense and sparse kernels based on profiling/metrics (e.g., non-zero ratio).
Toggling can be beneficial: sparse kernels reduce wasted work when inputs are highly sparse, while dense kernels can be faster when inputs are closer to dense due to better locality and vectorization.
The rewrite preserves semantics while improving performance across queries where sparsity varies by dataset.

\paragraph{Relationalizing ML computation into join/aggregate form.}
\texttt{Mul2Join}-\texttt{AggRewriteAction} rewrites matrix multiplication computations (e.g., a dot product $\sum_k x_k\!\cdot\! w_k$) into an equivalent \emph{cross join + aggregate} plan by materializing feature/value pairs and weights as relations, pairing them by feature index, and aggregating products with \texttt{SUM}.
\texttt{DecisionForestUDF2RelationRewriteAction} similarly replaces a decision-forest inference UDF with a \emph{cross join + aggregate} formulation that computes per-tree outputs and aggregates them to obtain the final prediction.
These rewrites expose ML inference as standard relational operators, enabling classic optimizations such as join reordering, pushdown, and vectorized execution while preserving semantics.

\paragraph{Fusing neural-network UDF chains into Torch operators}
These actions target neural-network inference written as nested scalar UDF calls (often one UDF per layer or per sub-expression).
\texttt{TwoLayerUDF2}-\texttt{TorchNNRewriteAction} handles a common two-layer pattern, while \texttt{MultiLayerUDF2Torch}-\texttt{NNRewriteAction} generalizes to deeper networks; both replace the UDF chain with a single fused Torch NN operator that represents the full model.
This reduces per-tuple interpretation overhead, improves vectorized execution, and gives the optimizer a clearer cost boundary for planning.
\texttt{MultiLayerUDF2}-\texttt{TorchNNCUDArewriteAction} further selects a CUDA-backed implementation of the fused operator when GPU support is available and beneficial, accelerating inference while preserving semantics.

\paragraph{ML decomposition and fine-grained pushdown}
\texttt{MLDecomposition}-\texttt{PushdownRewriteAction} decomposes a compound ML inference expression into smaller sub-operators and pushes them down the plan when safe (e.g., across projections and towards filters/joins), so expensive inference runs on fewer rows by leveraging selectivity and join-cardinality behavior.
\texttt{MLFactorizationRewriteAction} further enables pushdown at a finer granularity for linear/affine forms (e.g., \texttt{XW + b}) by factorizing the computation across joined sources: when $X = [X_1 \mid X_2 \mid \cdots]$, it splits $W = [W_1; W_2; \cdots]$ and rewrites $XW + b = X_1W_1 + X_2W_2 + \cdots + b$.
Together, these rewrites move ML work closer to where features are produced, reducing post-join computation and intermediate data movement while preserving semantics.

\paragraph{Reducing tree inference cost using constraints (\texttt{TreeModelPruning}-\texttt{RewriteAction}).}
This action applies to tree-based models and uses constraints already present in the query (e.g., filter predicates or known feature bounds) to identify branches that can never be taken.
When pruning is semantics-preserving, it removes unreachable branches from evaluation.
This reduces the amount of decision logic executed per tuple and speeds up inference without changing results.

% --- action library table (fixed-width columns so it doesn't overflow) -----
\begin{table}[h]
\centering
\small
\caption{Built-in rewrite actions in the \OptBench demo action library.}
\setlength{\tabcolsep}{3pt}
\renewcommand{\arraystretch}{1.12}

% helper: typeset long identifiers with good breakpoints

\begin{tabularx}{\columnwidth}{
  >{\raggedright\arraybackslash}p{0.4\columnwidth}
  >{\raggedright\arraybackslash}X
}
\toprule
\textbf{Rewrite Action} & \textbf{Purpose (inference rewrite)} \\
\midrule
MatMulDense2Sparse &
Switch/annotate matrix multiplication to a sparse variant when sparsity metrics indicate benefit. \\

DecisionForestUDF2Relation &
Rewrite decision-forest inference UDFs into an equivalent relational form to enable pushdown and reuse. \\

MatMul2Relation &
Rewrite eligible multiply/aggregation patterns into a join+aggregate form to improve execution structure. \\

ConvNN2MatMul &
Rewrite the convolutional operator as a matrix multiplication of the flattened image and concatenated filters.\\

MultiLayerUDF2TorchNN &
Replace a multi-layer NN UDF expression with a fused neural-network operator. \\

MLDecompositionPushdown &
Decompose compound ML inference expressions and push computation closer to feature sources when safe. \\

Fuse2TorchNN &
Specialized rewrite for two-layer NN UDF patterns into a fused NN operator. \\

MLFactorization &
Factor shared ML subexpressions/weights to reduce redundant inference work. \\

TreeModelPruning &
Prune redundant parts of tree models (when safe) to reduce inference cost. \\
\bottomrule
\end{tabularx}

\label{tab:action-library}
\end{table}

\paragraph{Customization of Rewrite Actions.}
Importantly, actions are extensible via a simple registration interface, allowing
researchers to easily develop and plug-in new rewrite actions.

Algorithm~\ref{alg:action-apply-plan-rewrite} shows the generic method for applying $A$ an action to rewrite a query plan stored as a tree structure rooted at $p$ through tree traverse.
During the iterative tree traversal process, at each iteration, it will  
(1) see whether the current plan node is eligible for applying $A$ via $A$'s \textsc{MatchesPlan} method, if it matches, we will apply \textsc{RewritePlan}; (2) it will then traverse the expression tree at this node (e.g., the expression tree representing a filter predicate customizing a selection operator or an expression tree that customize a projection operator), and for each expression, it applies  $A$'s \textsc{RewriteExpr} method; and (3) it moves to the next node in the query plan tree.
The algorithm returns both the rewritten plan and a boolean flag indicating whether any modification occurred, enabling optimizers to apply actions
iteratively (e.g., until reaching a fixpoint or exhausting a search budget).

\begin{algorithm}[t]
\caption{\textsc{ActionApplyPlanRewrite}$(A, p)$}
\label{alg:action-apply-plan-rewrite}
\KwIn{Action $A$, query plan tree node $p$}
\KwOut{Tuple $(modified, p')$ where $modified \in \{\texttt{true}, \texttt{false}\}$}

$modified \leftarrow \texttt{false}$\;

% (1) Operator-level rewrite (optional)
\If{$A.\textsc{MatchesPlan}(p)$}{
    $p \leftarrow A.\textsc{RewritePlan}(p)$\;
    $modified \leftarrow \texttt{true}$\;
}

% (2) Expression-level rewrite within this plan node (optional)
\ForEach{expression $e$ contained in $p$}{
    $(m_e, e') \leftarrow A.\textsc{RewriteExpr}(e)$\;
    Replace $e$ with $e'$ in $p$\;
    $modified \leftarrow modified \lor m_e$\;
}

% (3) Recurse into child plan nodes
\ForEach{child plan node $c$ of $p$}{
    $(m_c, c') \leftarrow \textsc{ActionApplyPlanRewrite}(A, c)$\;
    Replace $c$ with $c'$ in $p$\;
    $modified \leftarrow modified \lor m_c$\;
}

\Return $(modified, p)$\;
\end{algorithm}

Taking the \texttt{MatMulDense2SparseRewriteAction} as an example, its \textsc{RewriteExpr} implements a pattern match for a \texttt{matrix\_multiply} operator with dense input (the density of the input data being greater than a tunable threshold); when
matched, it rewrites the operator by switching its execution mode/annotation to a sparse variant while preserving semantics.
%The same action template supports inference-centric operator rewrites such as decomposing compound ML UDFs, pushing inference computation closer to feature
%sources, selecting specialized kernels (e.g., GPU-backed variants), or rewriting model inference into relational form, provided the rewrite is
%semantics-preserving (or explicitly marked as approximate).

\subsection{Extensible Statistics Estimation}
\label{sec:metric-estimation}
\OptBench provides a library of statistics estimation methods  that can be shared across optimizers to facilitate optimizer development. Various types of statistics are
derived from (1) data statistics (e.g., the sparsity, cardinality, and arity of relations or intermediate results), (2) predicate selectivity and the ratio of join input/output cardinalities (denoted as join ratio), and (3) time and space complexity of ML operators. We use DuckDB’s native cardinality estimation as the baseline estimate for
relational operators, and augment it with targeted profiling that computes ML-relevant statistics (e.g., sparsity and structural zeros, and ML operators' parameter size) that are typically
absent from relational catalogs. Examples are illustrated in Tab.~\ref{tab:metric-catalog}.

To estimate ML-oriented metrics without executing the full query, \OptBench issues small sampling queries over the feature inputs referenced by the plan.
These samples are used to compute robust summary statistics that guide inference rewrites such as switching dense matrix multiplication to sparse kernels or
delaying expensive inference until after selective operators. Sampling keeps metric collection lightweight while remaining sensitive to query- and
data-dependent properties.

\begin{table}[t]
\centering
\small
\caption{Examples of supported statistics in \OptBench.}
\setlength{\tabcolsep}{4pt}
\renewcommand{\arraystretch}{1.15}
\begin{tabularx}{\columnwidth}{l X}
\toprule
\textbf{Statistics} & \textbf{How computed} \\
\midrule
cardinality / selectivity estimation & DuckDB estimator (baseline) \\
nnz\_ratio & total non-zero elements\\
zero\_rows, zero\_cols & fraction of fully-zero rows/cols \\
ML model FLOPS & via ML operator's derived get\_shape() method \\
num\_parameters & via ML operator' derived get\_shape() method \\
forest\_num\_trees & Forest model metadata \\
\bottomrule
\end{tabularx}
\label{tab:metric-catalog}
\end{table}

\subsection{Diverse and Extensible Inference Queries}
\label{sec:queries}

\OptBench includes ten SQL+ML inference queries drawn from two widely used sources (Tab. \ref{tab:query-suite-summary}).
The first three queries are adapted from Raven\cite{park2022end}'s prediction-query workloads, which are evaluated on real-world datasets commonly used in data science tasks, including the \textsc{Expedia} and \textsc{Flights} relational datasets from Project Hamlet and a \textsc{CreditCard} fraud dataset from Kaggle.
These workloads are intentionally join-heavy (e.g., \textsc{Expedia} is a 3-way join and \textsc{Flights} is a 4-way join) and feature high-dimensional, mostly one-hot encoded categorical inputs, making them representative stress tests for optimizer behavior around joins, sparsity, and inference placement.

The next five queries are based on selected TPCx-AI use cases, a vendor-neutral TPC standard benchmark that specifies end-to-end AI pipelines and a suite of ten business-oriented tasks.
We include (UC01) customer segmentation (clustering), (UC03) store sales forecasting, (UC04) review spam detection (text classification), (UC08) multiclass classification over operational records, and (UC10) online fraud/risk scoring.
Together, these eight queries cover both structured/tabular and text modalities and span a range of model families (trees/ensembles, boosting, clustering, Naive Bayes, and DNN inference), providing a compact but diverse target set for benchmarking SQL+ML rewrite actions and optimizers.

Finally, we include two queries derived from IDNet~\cite{xie2024idnet} to capture \emph{image-centric} inference pipelines inside SQL. 
\textsc{Q\_IDNet1} joins identity-document images with structured toll-audit records on a shared identifier (e.g., license number) and applies a CNN inference UDF as a selective filter to retain only non-fraud cases; this query stresses predicate placement, join ordering, and the cost impact of running expensive vision inference before or after relational operators. 
\textsc{Q\_IDNet2} models an LLM-based fraud detector where each input image is judged multiple times using different retrieved reference examples (implemented as a Cartesian join over demonstration pairs), followed by a majority-vote aggregation to produce the final label; this query highlights repeated black-box inference, aggregation-heavy post-processing, and optimizer trade-offs around reordering, caching/materialization, and pushing down selective predicates.
Together, these two queries extend the suite beyond tabular and text workloads by introducing multimodal inference patterns that are increasingly common in modern AI-enabled data systems.

%\jia{Jay: Please describe the set of queries supported in \OptBench in more detail here. For example, what are the datasets involved. How popular are those datasets/benchmarks? How many different types of modalities are used? How many Why do you select those queries? DONE}

% ---- Table ----
\begin{table}[t]
\centering
\caption{Summary of the benchmark query suite: dataset, task (what the query returns), and the inference model invoked. }
\small
\setlength{\tabcolsep}{4pt}
\renewcommand{\arraystretch}{1.15}

\begin{tabular}{
  >{\raggedright\arraybackslash}p{0.13\columnwidth}
  >{\raggedright\arraybackslash}p{0.14\columnwidth}
  >{\raggedright\arraybackslash}p{0.5\columnwidth}
  >{\raggedright\arraybackslash}p{0.17\columnwidth}
}
\toprule
\textbf{Query} & \textbf{Dataset} & \textbf{Task / Output} & \textbf{ML UDF} \\
\midrule
Q\_Expedia & Expedia~\cite{project_hamlet} &
Hotel search ranking: returns a model score per candidate listing within a search instance (per \texttt{(srch\_id, prop\_id)}) after selective filters. &
Decision Tree \\

Q\_Flights & Flights~\cite{project_hamlet} &
Route attribute prediction: returns a predicted label (e.g., codeshare/route property) per route after joining airline + source/destination airport metadata. &
Decision Forest \\

Q\_Credit & Credit-Card~\cite{kaggle_creditcardfraud} &
Fraud detection: returns a predicted risk/class per transaction from \texttt{V1..V28} and \texttt{Amount}. &
XGBoost \\

Q\_UC01 & TPCx-AI~\cite{tpcx-ai} &
Customer clustering: returns a cluster assignment per customer using return-ratio and purchase-frequency features (with min--max scaling). &
KMeans++ \\

Q\_UC03 & TPCx-AI &
Forecasting-style inference: returns a prediction per \texttt{(store, department)} from encoded categorical features and normalized week index. &
DNN \\

Q\_UC04 & TPCx-AI &
Spam classification: returns \texttt{predicted\_spam} per review document using tokenization/stemming + Naive Bayes scoring. &
Naive Bayes \\

Q\_UC08 & TPCx-AI &
Order classification: returns a multiclass prediction per order from aggregated lineitem/department + weekday features. &
DNN + Softmax \\

Q\_UC10 & TPCx-AI &
Fraud/risk scoring: returns a sigmoid probability per transaction from normalized amount and business-hour features. &
DNN + Sigmoid \\

Q\_IDNet1 & IDNet~\cite{xie2024idnet} &
Join + CNN filter: keep only non-fraud images after linking to toll-audit data. 
 &
CNN \\

Q\_IDNet2 & IDNet &
LLM voting: repeatedly judge fraud for each input image using references, then majority-vote the final label.
 &
LLM \\

%Q\_IDNet3 & IDNet &
%\textcolor{blue}{Doug TODO} &
%GPT-Image \\

\bottomrule
\end{tabular}
\label{tab:query-suite-summary}
\end{table}

\subsection{Web Interfaces}

%\jia{Jay: Please describe the functionalities supported by the web interface here.}

\begin{figure*}[t]
    \centering
    \includegraphics[width=\linewidth]{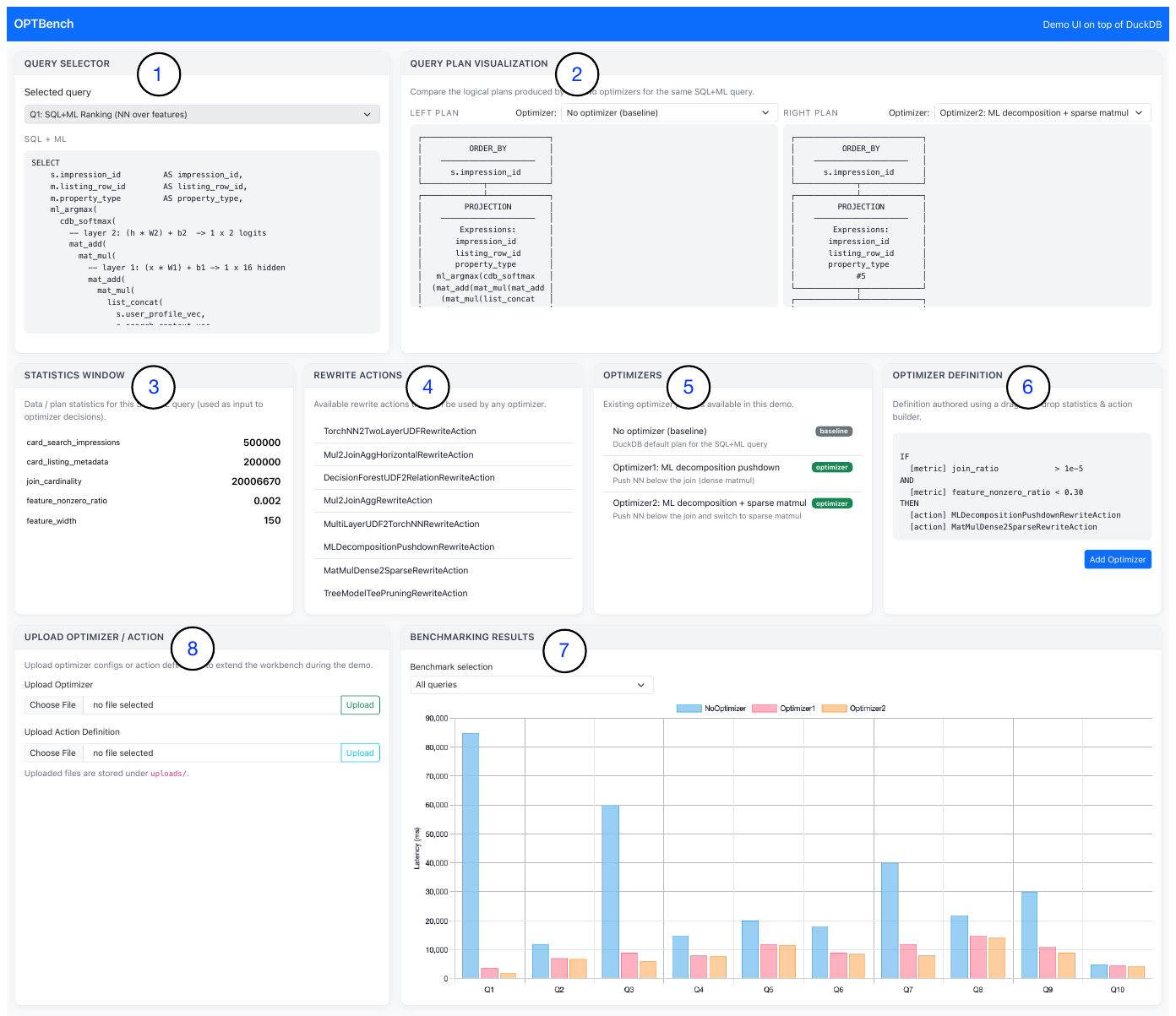}
    \caption{OptBench workbench overview. Users select a SQL+ML query (top-left), compare two optimizers side-by-side (top-center), and inspect the statistics, rewrite action
    catalog, optimizer profiles, and rule definitions (middle). The workbench also supports uploading external optimizer/action definitions
    (bottom-left) and benchmarking latency across the query suite (bottom-right).}
    \label{fig:demo-workbench}
\end{figure*}

\OptBench exposes a web-based optimizer workbench (Fig.~\ref{fig:demo-workbench}) organized into eight panels: (1) a \emph{Query Selector} that chooses a SQL+ML query from the suite, (2) a \emph{Query Plan Visualization} view that renders logical plans side-by-side for direct comparison, (3) a \emph{Statistics Window} that displays the statistics computed for the selected query, (4) an \emph{Rewrite Actions Window} that lists available rewrite actions, (5) an \emph{Optimizers} panel that lists built-in and user-defined optimizer profiles, (6) an \emph{Optimizer Definition} panel that lets users compose rule-based optimizers by mapping statistics to actions, (7) a \emph{Benchmarking Results} view that summarizes end-to-end latency across the query suite, and (8) an \emph{Upload Optimizer/Action} panel for importing external optimizer profiles and action definitions into the workbench.
 %\jia{Jay, please mark the numbers correspondingly in Fig.~\ref{fig:demo-workbench} DONE}.

\section{Use Cases}
In this section, we will introduce two use cases of \OptBench: Optimizer Development and Performance Comparison. 
\subsection{Optimizer Development, Testing, and Evaluation}
\label{sec:optimizer-design}
\OptBench facilitates the development of rule-based or cost-based SQL-ML query optimizers by providing a library of rewrite actions and statistics. In addition, \OptBench also implements a set of queries and provides query plan visualization through its web interface, which will facilitate the testing, evaluation, and debugging of the query optimizer.

\paragraph{Optimizer profile registration.}
As illustrated in Listing.~\ref{list:registration}, \OptBench exposes a lightweight registration interface so that each optimizer is treated as a named \emph{optimizer profile} that can be activated and listed in the web-based user interface (UI) for selection for benchmark comparisons. Internally, a profile binds an optimizer identifier to an optimizer entry point, which receives the
initial DuckDB query logical plan and the query context (including a list of statistics) and returns an optimized logical plan structure. %This design decouples the UI and benchmarking pipeline independent of the optimizer implementation.

\begin{lstlisting}[language=C++, caption={Registering an optimizer profile in OptiBench (illustrative) \label{list:registration}}, label={lst:optimizer-register}]
struct OptimizerProfile {
  std::string name;
  std::function<void(OptimizerExtensionInput&,
                     std::unique_ptr<LogicalOperator>&)> optimize_fn;
};

void RegisterOptimizer(OptimizerProfile p);

RegisterOptimizer({
  /*name=*/"DP-CostOpt",
  /*optimize_fn=*/DPCostOpt::Run
});
\end{lstlisting}

\paragraph{Example 1: Developing rule-based optimizers.}
A rule-based optimizer can be easily developed in \OptBench by specifying a prioritized set of rules that map query- and data-dependent statistics to semantics-preserving inference
rewrite actions. Each rule is a predicate over the statistics or conditions regarding data, ML operators, expressions, or query plan, and when it fires, it triggers one or more actions
from the library of registered rewrite actions. Rewrite actions are applied to the current logical plan following Algorithm~\ref{alg:action-apply-plan-rewrite}. \OptBench records a decision trace containing fired rules and the structural deltas induced by each action,
enabling users to inspect how domain knowledge translates into concrete plan rewrites and how
those rewrites affect latency under unified benchmarking.
%\jia{Jay: Please add an algorithm to illustrate the pseudo code snippet for a rule-based optimizer for a longer version. I remember you had such example before. Then, please reference it in the text. DONE}

\begin{figure}[t]
    \centering
    \includegraphics[width=\columnwidth]{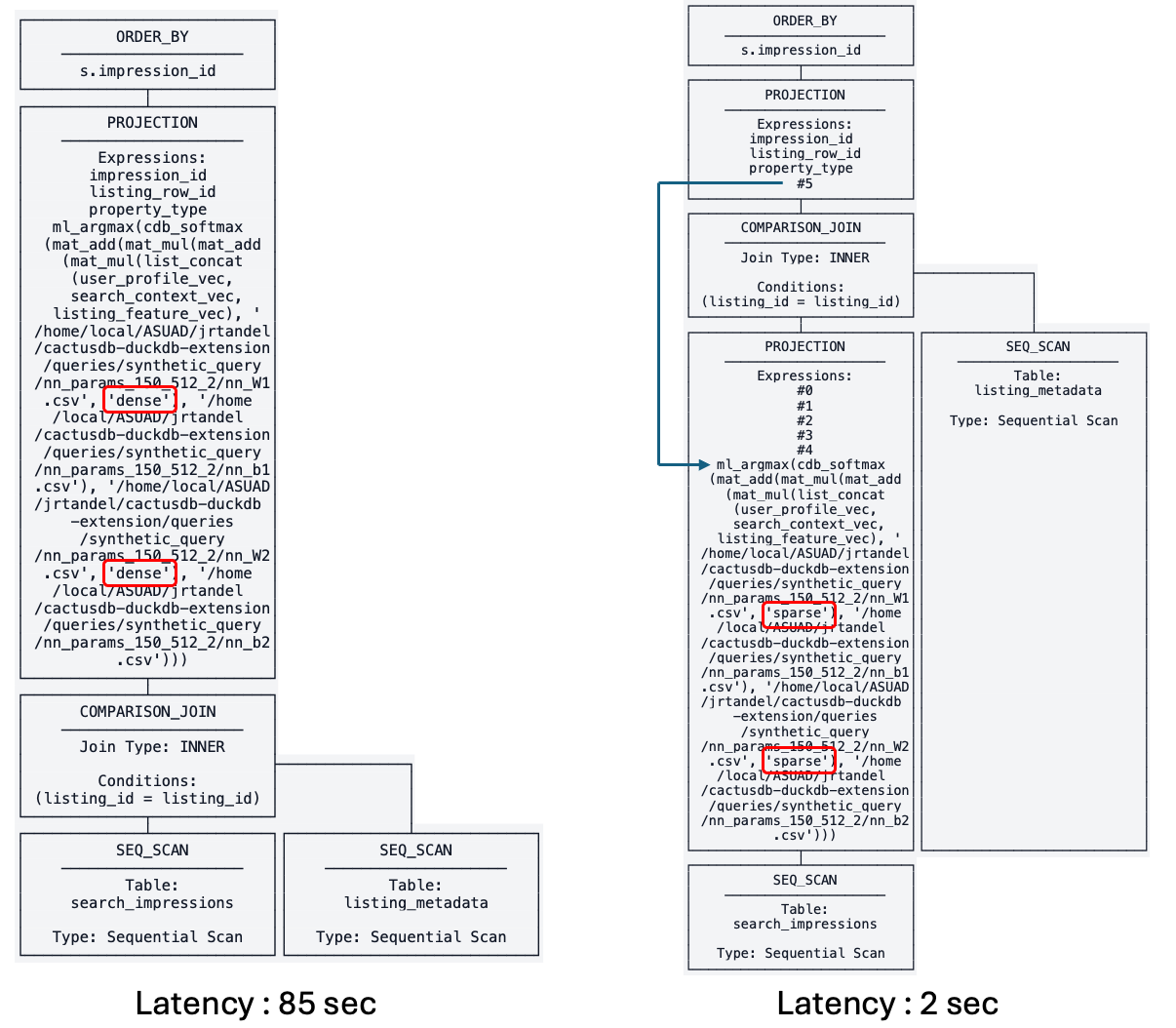}
    \caption{Effect of the example rule-based optimizer on an inference query. Starting from the baseline plan (no optimizer), the optimizer fires a metric-driven rule that invokes the actions highlighted in the figure (e.g., \texttt{MLDecompositionPushdownRewriteAction} and \texttt{MatMulDense2SparseRewriteAction}).}
    \label{fig:opt-effect}
\end{figure}

Figure~\ref{fig:opt-effect} shows how the example OptiBench optimizer transforms the baseline logical plan (Plan A, no optimizer) into an optimized plan (Plan B) by applying the inference rewrites indicated in the figure. 
The optimizer is specified as a simple metric-driven rule: when the query exhibits a large intermediate join result and sufficiently sparse feature vectors, it triggers MLDecompositionPushdownRewriteAction followed by MatMulDense2SparseRewriteAction. 
At runtime, OptiBench evaluates the rule predicate over the computed metric vector and materializes it into a small conditional code block that invokes the registered actions using Algorithm~\ref{alg:action-apply-plan-rewrite}. 
The resulting plan pushes NN inference below the join (so fewer tuples are scored) and switches dense \texttt{mat\_mul} operators to sparse variants (so each scored tuple is cheaper). 
Together, these changes reduce end-to-end latency from $85\,\text{s}$ to $1.976\,\text{s}$ as shown in the figure.

\paragraph{Example 2: Developing cost-based optimizers.}
A cost-based optimizer in \OptBench selects an optimized inference plan by \emph{enumerating} alternative plans induced by the action library and choosing
the lowest-cost candidate under a cost model and a searching strategy. In addition to the rewrite actions, \OptBench's library of statistics methods would greatly facilitate the cost model development. However, the programmers must define the cost function and develop the search strategy. %Concretely, starting from the initial plan, the optimizer repeatedly applies all applicable inference actions (e.g., decomposition pushdown, kernel mode switches, safe pushdowns) up to a bounded budget, memoizes the best-scoring representative per canonical plan state, and returns the minimum-cost plan found. This decouples \emph{plan generation} (via reusable actions) from \emph{plan evaluation} (via the scoring function), allowing OptiBench to compare rule-driven and cost-driven optimization on the same inference workloads using the same plan representation, action library, and benchmarking workflow.
%
%\paragraph{Illustrative implementation (cost-based DP).}
Listing~\ref{lst:dp-cost-opt-compact} sketches a compact C++-style implementation of the cost-based optimizer as a depth-bounded DP enumerator using \OptBench.

\begin{lstlisting}[language=C++, caption={Compact cost-based DP optimizer (enumerate inference rewrite actions)}, label={lst:dp-cost-opt-compact}]
static void Run(OptimizerExtensionInput &in,
                std::unique_ptr<LogicalOperator> &plan) {
  int L = 2; // rewrite-depth budget
  // actions: inference rewrite actions (DecomposePushdown, Dense2Sparse, ...)
  // Score(p): cost model using (in.metrics, plan features)
  // Hash(p): canonical plan hash (memoization)

  auto best = plan->Copy();
  double best_cost = Score(best, in);

  std::vector<std::unique_ptr<LogicalOperator>> frontier{ plan->Copy() };
  std::unordered_map<size_t,double> memo; memo[Hash(frontier[0])] = best_cost;

  for (int d = 0; d < L; d++) {
    std::vector<std::unique_ptr<LogicalOperator>> next;
    for (auto &p : frontier) {
      for (auto &a : actions) {                 // expand all actions
        if (!a.Applicable(p, in)) continue;
        auto p2 = p->Copy(); a.Apply(p2, in);
        double c = Score(p2, in); size_t h = Hash(p2);
        if (!memo.count(h) || c < memo[h]) {    // keep best per state
          memo[h] = c; next.push_back(p2->Copy());
          if (c < best_cost) { best_cost = c; best = p2->Copy(); }
        }
      }
    }
    if (next.empty()) break;
    frontier = std::move(next);
  }
  plan = std::move(best);
}
\end{lstlisting}

\subsection{Performance Comparison and Benchmarking}
\OptBench benchmarks each optimizer on a shared and extensible suite of \emph{SQL+ML multimodal inference queries} as described in Sec.\ref{sec:queries}. All registered optimizers, including DuckDB's default optimizer, our built-in rule-based and dynamic programming-based SQL-ML query optimizers, and user-developed optimizers,
%the baseline optimizer, rule-based OptiBench optimizers, and plug-in backends (e.g., search-based) 
are evaluated using the \emph{same execution backend}, the same queries, and the same data, ensuring a unified measurement environment and enabling \emph{apples-to-apples} comparison.

\OptBench reports two views. First, a \emph{plan comparison} view highlights structural differences between optimized plans (e.g., operator placement and
rewrite-induced changes), making it easier to attribute performance differences to specific inference rewrites. Second, a \emph{benchmarking panel}
summarizes end-to-end latency (and other runtime metrics) across the shared query suite for all selected optimizers under the same backend and data.
Together, these views support an apples-to-apples benchmarking workflow that connects optimizer choices to both plan-level changes and measurable
performance outcomes.

\section{Interactive Demonstration Design}
In our proposed demonstration, users experience the \OptBench capabilities  by (i) interactively develop a rule-based optimizer by defining rules that trigger rewrite actions based on query- and data-dependent statistics using our web-based interface, including testing, debugging, and performance evaluation; (ii) register and upload our predefined rule-based and cost-based optimizers through the web-based interface; and (iii) benchmark multiple optimizers on one or more user-selected SQL+ML queries for \emph{transparent, apples-to-apples comparison}: all optimizers are evaluated on the same query suite and executed on the same backend system (our DuckDB-based prototype), ensuring that differences in latency and plan structure are attributable to optimizer design decisions rather than implementation artifacts of the underlying systems. These interactive scenarios are detailed as follows.

\subsection{Scenario 1. Developing a Custom Optimizer using \OptBench}
Every demonstration participant can use their database system knowledge to quickly develop, test, debug, and evaluate a simple rule-based SQL-ML query optimizer using \OptBench. Using \OptBench, the entire process may only take a few minutes for an optimizer that has only one or two rules. For example, A participant wants to test a hypothesis such as: ``for large joins and sparse feature vectors, it is beneficial to push ML computation downward and use sparse kernels.'',  
%A query optimizer first inspect the sparsity of data input to each linear algebra operator using the corresponding statistics estimation method, and then decide whether the current query instance satisfies the preconditions (e.g., join ratio above a threshold and feature nonzero ratio below a sparsity threshold), and then include 
s/he could define a new rule such as "if a dense \texttt{matrix\_multiply} operator's input data has more than $1,000,000$ rows and the sparsity is above $0.7$, the rewrite action named \texttt{MatMulDense2SparseRewriteAction} would be triggered to invoke the corresponding sparse \texttt{matrix\_multiply}  operator instead. Then the participant can code that rule to compose a simple query optimizer in the \emph{{Optimizer Definition} \circled{6}} panel. 
Once saved, this new optimizer profile appears alongside the other optimizers registered and listed in the \emph{Optimizers} panel and benchmarking views. The user can then place any two optimizers side by side in the plan comparison pane and inspect their latency bars in the \emph{Benchmarking Results} panel, turning high-level hypotheses about metric-driven optimization into concrete, measurable experiments. 
In a small pilot with a new user, implementing the rule above (end-to-end, from opening the workbench to saving the optimizer) took 4 minutes.
%\textcolor{red}{Needs a simple user study: The end-to-end process only requires less than ten minutes for a new user.} 
%\jia{Jay: for each panel please give a number like \circle{3} corresponding to the number in Fig~\ref{fig:demo-workbench} after the name, e.g., \emph{Benchmarking Results \circled{6}} DONE}.

\begin{comment}
In addition, for rule-based optimizers that are developed following our defined conditional flow template, \OptBench can visualize the mapping from rule definitions to an executable optimizer profile (Fig.~\ref{fig:rules-to-profile}), for easy understanding. %making the rule-to-action intent explicit and reproducible.

\begin{figure}[t]
    \centering
    \includegraphics[width=\linewidth]{images/rules_to_profile.png}
    \caption{Rule-based optimizer design. Users define an optimizer profile as metric predicates that trigger actions (left), which OptiBench materializes
    as an executable optimizer configuration/profile (right).\textcolor{red}{The text font size is too small. I can barely read it. Please reorganize the figure to make it readable!}}
    \label{fig:rules-to-profile}
\end{figure}
\end{comment}

\subsection{Scenario 2. Comparing Optimizers on a SQL+ML Query}
A participant starts by selecting an SQL+ML query from the \emph{Query Selector} (Fig.~\ref{fig:demo-workbench}). %The \emph{\textcolor{red}{Statistics Panel}} immediately reveals key properties of the query instance, such as base table cardinalities, join cardinality/ratio, and feature statistics (e.g., sparsity and feature width). 
The participant then chooses two optimizers.
For example, the user may select a DuckDB's default optimizer and our developed rule-based or cost-based SQL-ML query optimizer.%on the left and a domain-aware optimizer on the right.
The baseline plan may keep the ML operator as a single high-level expression above a join with large output cardinality, while the SQL-ML query optimizer applies a sequence of rewrite actions (e.g., decomposing an ML operator and pushing computation closer to the feature sources). The side-by-side plan visualization view makes structural differences explicit, while the \emph{Benchmarking Results} panel (Fig.~\ref{fig:demo-workbench}) quantifies their end-to-end latency on the same backend and dataset. This scenario demonstrates how \OptBench supports transparent comparisons between heterogeneous optimizers on ML-heavy workloads. 

\subsection{Scenario 3. Extending \OptBench}
Cost-based or complex rule-based optimizers are difficult to express purely as rules (e.g., complex search strategies or specialized heuristics) and would be hard to develop in the``{Optimizer Definition \circled{6}}'' panel. To support these cases, \OptBench provides an extension path for registering external optimizers, and required rewrite action definitions through the backend optimizer interface. In the demo, users can upload an optimizer profile or an action definition via the \emph{Upload Optimizer/Action} view. After registration, the external optimizer appears in the \emph{Optimizers} panel and can be compared against built-in and rule-based optimizers using the same plan comparison and benchmarking workflow.

This scenario highlights that \OptBench is not limited to a single optimizer family: it provides a unified evaluation loop where both user-authored, rule-based optimizers and externally implemented optimizers can be inspected (via plan diffs) and compared quantitatively (via latency over the query suite).

A typical demonstration workflow may also follow three steps:
(1) select a query and inspect its relevant statistics (you can also select multiple queries), (2) choose two or more optimizers to compare their plans and latency,
and (3) optionally create a new optimizer profile (via rules and actions) or register an external optimizer (we will prepare a few such optimizers for demonstrating this step), then re-run benchmarking to quantify the impact of the new design.

\bibliographystyle{ACM-Reference-Format}
\bibliography{jia}

\appendix
%\newpage
\appendix
\section{Benchmark Query Suite}
\label{app:query-suite}

\begin{lstlisting}[language=SQL, caption={Q\_Expedia (decision tree inference)}, label={lst:q-expedia}]
SELECT
  Expedia_S_listings_extension2.prop_id,
  Expedia_S_listings_extension2.srch_id,
  udf(prop_location_score1, prop_location_score2, prop_log_historical_price, price_usd,
      orig_destination_distance, prop_review_score, avg_bookings_usd, stdev_bookings_usd,
      position, prop_country_id, prop_starrating, prop_brand_bool, count_clicks, count_bookings,
      year, month, weekofyear, time, site_id, visitor_location_country_id, srch_destination_id,
      srch_length_of_stay, srch_booking_window, srch_adults_count, srch_children_count,
      srch_room_count, srch_saturday_night_bool, random_bool)
FROM Expedia_S_listings_extension2
JOIN Expedia_R1_hotels2
  ON Expedia_S_listings_extension2.prop_id = Expedia_R1_hotels2.prop_id
JOIN Expedia_R2_searches
  ON Expedia_S_listings_extension2.srch_id = Expedia_R2_searches.srch_id
WHERE prop_location_score1 > 1
  AND prop_location_score2 > 0.1
  AND prop_log_historical_price > 4
  AND count_bookings > 5
  AND srch_booking_window > 10
  AND srch_length_of_stay > 1;
\end{lstlisting}

\begin{lstlisting}[language=SQL, caption={Q\_Flights (decision forest inference)}, label={lst:q-flights}]
SELECT
  Flights_S_routes_extension2.airlineid,
  Flights_S_routes_extension2.sairportid,
  Flights_S_routes_extension2.dairportid,
  udf(slatitude, slongitude, dlatitude, dlongitude,
      name1, name2, name4, acountry, active,
      scity, scountry, stimezone, sdst,
      dcity, dcountry, dtimezone, ddst) AS codeshare
FROM Flights_S_routes_extension2
JOIN Flights_R1_airlines2
  ON Flights_S_routes_extension2.airlineid = Flights_R1_airlines2.airlineid
JOIN Flights_R2_sairports
  ON Flights_S_routes_extension2.sairportid = Flights_R2_sairports.sairportid
JOIN Flights_R3_dairports
  ON Flights_S_routes_extension2.dairportid = Flights_R3_dairports.dairportid
WHERE name2 = 't'
  AND name4 = 't'
  AND name1 > 2.8;
\end{lstlisting}

\begin{lstlisting}[language=SQL, caption={Q\_Credit (XGBoost inference)}, label={lst:q-credit}]
SELECT
  Time,
  Amount,
  udf(V1, V2, V3, V4, V5, V6, V7, V8, V9, V10, V11, V12, V13, V14, V15,
      V16, V17, V18, V19, V20, V21, V22, V23, V24, V25, V26, V27, V28, Amount) AS Class
FROM Credit_Card_extension
WHERE V1 > 1
  AND V2 < 0.27
  AND V3 > 0.3;
\end{lstlisting}

%tpcxai

\begin{lstlisting}[language=SQL, caption={Q\_UC01 (KMeans++ inference with min--max scaling)}, label={lst:q-uc01}]
WITH table_groups AS (
  SELECT
    MIN(EXTRACT(YEAR FROM CAST(date AS DATE))) AS invoice_year,
    SUM(quantity * price) AS row_price,
    SUM(COALESCE(or_return_quantity, 0) * price) AS return_row_price,
    o_customer_sk AS customer_id,
    o_order_id    AS order_id
  FROM lineitem li
  LEFT JOIN order_returns oret
    ON li.li_order_id  = oret.or_order_id
   AND li.li_product_id = oret.or_product_id
  JOIN "order" ord
    ON li.li_order_id = ord.o_order_id
  GROUP BY o_customer_sk, o_order_id
),
table_ratios AS (
  SELECT
    customer_id,
    AVG(return_row_price / NULLIF(row_price, 0)) AS return_ratio
  FROM table_groups
  GROUP BY customer_id
),
table_frequency_groups AS (
  SELECT
    customer_id,
    invoice_year,
    COUNT(DISTINCT order_id) AS orders_per_year
  FROM table_groups
  GROUP BY customer_id, invoice_year
),
table_frequency AS (
  SELECT
    customer_id,
    AVG(orders_per_year) AS frequency
  FROM table_frequency_groups
  GROUP BY customer_id
),
features_raw AS (
  SELECT
    r.customer_id,
    f.frequency,
    r.return_ratio
  FROM table_ratios r
  JOIN table_frequency f
    ON r.customer_id = f.customer_id
),
features AS (
  SELECT
    customer_id,
    -- abstract min--max scaling (your min_max_scaler UDF)
    minmax_apply('uc01_scaler', ARRAY[frequency, return_ratio]) AS features
  FROM features_raw
)
SELECT
  customer_id,
  -- abstract KMeans assignment (cluster id)
  kmeans_predict('uc01_kmeans_model', features) AS cluster_id
FROM features;
\end{lstlisting}

\begin{lstlisting}[language=SQL, caption={Q\_UC03 (DNN inference over encoded store--department features)}, label={lst:q-uc03}]

WITH base AS (
  SELECT
    store,
    department,
    num_of_week,
    store_id_encoder(CAST(store AS INTEGER))     AS store_id_encoded,
    department_encoder(department)              AS department_encoded,
    CAST(num_of_week / 156.0 AS REAL)          AS num_of_week_norm
  FROM store_dept
),
feat AS (
  SELECT
    store,
    department,
    num_of_week,
    CAST(concat(store_id_encoded, department_encoded, num_of_week_norm) AS REAL[]) AS features
  FROM base
)
SELECT
  store,
  department,
  num_of_week,
  dnn_predict('uc03_model', features) AS prediction
FROM feat;
\end{lstlisting}

\begin{lstlisting}[language=SQL, caption={Q\_UC08 (DNN inference over order features; softmax output)}, label={lst:q-uc08}]

WITH ord AS (
  SELECT o_order_id, CAST(date AS TIMESTAMP) AS ts
  FROM "order"
),
ord2 AS (
  SELECT o_order_id, ts, day_of_week(ts) AS weekday
  FROM ord
),
j1 AS (
  SELECT
    o.o_order_id, o.ts, o.weekday,
    li.li_product_id, li.quantity
  FROM ord2 o
  JOIN lineitem li ON o.o_order_id = li.li_order_id
),
j2 AS (
  SELECT
    j1.o_order_id, j1.ts, j1.weekday,
    j1.quantity, p.department
  FROM j1
  JOIN product p ON j1.li_product_id = p.p_product_id
),
agg AS (
  SELECT
    o_order_id,
    ts AS date,
    department,
    quantity,
    SUM(quantity) AS scan_count,
    MIN(weekday)  AS weekday
  FROM j2
  GROUP BY o_order_id, ts, department, quantity
),
feat AS (
  SELECT
    o_order_id,
    date,
    CAST(concat(quantity, scan_count, weekday, department_encoder(department)) AS REAL[]) AS features
  FROM agg
)
SELECT
  o_order_id,
  date,
  softmax(dnn_predict('uc08_dnn_model', features)) AS prediction
FROM feat;
\end{lstlisting}

\begin{lstlisting}[language=SQL, caption={Q\_UC10 (DNN inference for fraud detection; sigmoid output)}, label={lst:q-uc10}]
WITH tx AS (
  SELECT
    transaction_id,
    sender_id,
    CAST(hour(CAST(time AS TIMESTAMP)) AS DOUBLE) AS business_hour_norm,
    amount
  FROM financial_transactions
),
j AS (
  SELECT
    tx.transaction_id,
    tx.business_hour_norm,
    tx.amount,
    fa.transaction_limit
  FROM financial_account fa
  JOIN tx ON fa.fa_customer_sk = tx.sender_id
),
feat AS (
  SELECT
    transaction_id,
    CAST(concat(amount / transaction_limit, business_hour_norm) AS REAL[]) AS features
  FROM j
)
SELECT
  transaction_id,
  sigmoid(dnn_predict('uc10_dnn_model', features)) AS prediction
FROM feat;
\end{lstlisting}

\begin{lstlisting}[language=SQL, caption={Q\_UC04 (Naive Bayes spam inference over tokenized reviews)}, label={lst:q-uc04}]
WITH docs AS (
  SELECT
    r.id,
    -- Text preprocessing UDFs (abstracted): clean -> tokenize -> stem
    ARRAY(
      SELECT stem_token(t)                       -- e.g., madlib.stem_token(t)
      FROM UNNEST(tokenize(clean(lower(r.text)))) AS t
      WHERE t <> '' AND t !~ '^[0-9]+$'
    ) AS tokens
  FROM review r
),
base_priors AS (
  -- Base class priors from the labeled training set (can also be stored as constants)
  SELECT
    LN(SUM(CASE WHEN spam = 1 THEN 1 ELSE 0 END) * 1.0 / COUNT(*)) AS log_spam_prior,
    LN(SUM(CASE WHEN spam = 0 THEN 1 ELSE 0 END) * 1.0 / COUNT(*)) AS log_non_spam_prior
  FROM uc04_train_preprocessed
),
token_scores AS (
  SELECT
    d.id,
    SUM(LN(COALESCE(m.spam_prob,     1e-9))) AS log_spam_score,
    SUM(LN(COALESCE(m.non_spam_prob, 1e-9))) AS log_non_spam_score
  FROM docs d
  CROSS JOIN UNNEST(d.tokens) AS tok(token)
  LEFT JOIN uc04_model m
    ON m.token = tok.token
  GROUP BY d.id
)
SELECT
  s.id,
  CASE
    WHEN s.log_spam_score + p.log_spam_prior
       > s.log_non_spam_score + p.log_non_spam_prior
    THEN 1 ELSE 0
  END AS predicted_spam
FROM token_scores s
CROSS JOIN base_priors p
ORDER BY s.id;
\end{lstlisting}

\begin{lstlisting}[language=SQL, caption={Q\_IDNet (Toll audit join with CNN predicate filtering)}, label={lst:q-idnet}]
SELECT
  lst.*,
  jd.*
FROM idnet AS lst
JOIN toll_audit AS jd
  ON lst.license_number = jd.license_number
WHERE CAST(cnn_predict(lst.imageData) AS INTEGER) = 0;
\end{lstlisting}

\begin{lstlisting}[language=SQL, caption={Q\_IDNet\_LLM\_Nested (nested LLM calls; majority vote)}, label={lst:q-idnet-llm-nested}]
WITH
input_imgs AS (
  SELECT license_number, imageData
  FROM idnet10k
  USING SAMPLE 10
),
ref1 AS (
  SELECT imageData AS ref_image1
  FROM idnet10k
  USING SAMPLE 500
),
ref2 AS (
  SELECT imageData AS ref_image2
  FROM idnet10k
  USING SAMPLE 500
),
votes AS (
  SELECT
    i.license_number,
    llm(
      'Tell me whether the INPUT image is fraudulent or not using the reference image(s) and the LLM responses.',
      i.imageData,
      r1.ref_image1,
      llm('Is this image fraudulent? Reply 1 for fraud, 0 otherwise.', r1.ref_image1),
      r2.ref_image2,
      llm('Is this image fraudulent? Reply 1 for fraud, 0 otherwise.', r2.ref_image2)
    ) AS vote
  FROM input_imgs i
  CROSS JOIN ref1 r1
  CROSS JOIN ref2 r2
)
SELECT
  license_number,
  CASE WHEN AVG(CAST(vote AS DOUBLE)) >= 0.5 THEN 1 ELSE 0 END AS is_fraud,
  AVG(CAST(vote AS DOUBLE)) AS fraud_vote_ratio,
  COUNT(*) AS num_votes
FROM votes
GROUP BY license_number
ORDER BY fraud_vote_ratio DESC;
\end{lstlisting}

\end{document}